# European summer weather regimes 1990-2019: Automatic classification and representation in a small global climate model ensemble

Sibille Wehrmann[1,*], Carolyne Pickler[1], Marlene Schramm[2], Thomas Mölg[1]

[1] Climate System Research Group, Institute of Geography, Friedrich-Alexander-Universität, Erlangen-Nürnberg, Bavaria, Germany

[2] Department of Physics, Friedrich-Alexander-Universität, Erlangen-Nürnberg, Bavaria, Germany

**Correspondence:** Sibille Wehrmann, sibille.wehrmann@fau.de

## Graphical abstract

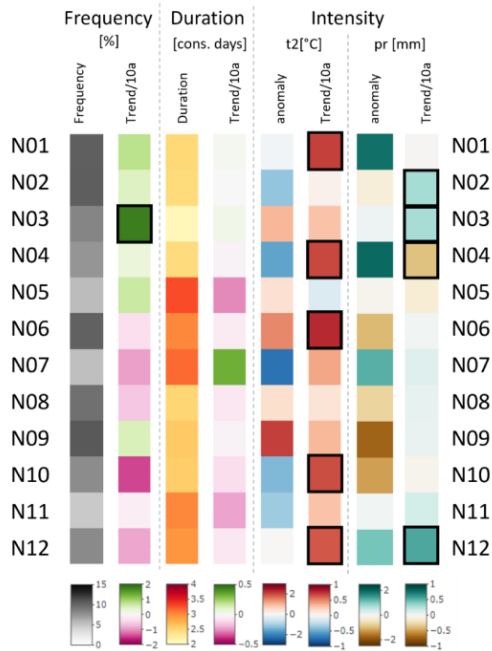

# Abstract

In Central Europe, the occurrence of different weather regimes (WRs) plays a major role in spatiotemporal temperature and precipitation patterns. In the context of increasingly extreme summers, this study focuses on European summer WRs (June-August, JJA) over the last three decades (1990-2019), and aims to examine the changing characteristics of these WRs and their potential implications. In addition, based on ERA5 reanalysis data, the WR representation from a carefully preselected, small ensemble of global general circulation models (GCMs) is analyzed. A methodological refinement concerns the combination of Self-Organizing Maps (SOM) with a novel GCM selection technique, which enhances the robustness of the simulated large-scale circulation patterns. WRs are defined using daily sea level pressure (SLP) and wind in the upper troposphere. Results reveal that the SOM captures predominant European summer synoptic patterns, and the salient result is a positive trend in 2 m air temperature across nearly all WRs. The selected GCMs – MPI-ESM1-2-LR r29i1p1f1, CanESM5 r1i1p1f1 and MRI-ESM2-0 r5i1p1f1 – identify WRs correctly and ERA5-based results are always within the range of this small GCM ensemble. Its members clearly show skill in representing large-scale WRs accurately and, thus, serve as valuable tools for studying synoptic weather patterns during summer in Central Europe. Therefore, we can recommend these GCMs for studies on WR-related climate projections of future summer conditions, particularly for those interested in climate impacts in Central Europe from a synoptic-scale perspective.

# 1. Introduction

The effects of climate change have been omnipresent in recent decades around the globe. A much-discussed topic in this context is the risk of extreme temperatures, drought and summer heatwave events, which are likely to increase their frequency and intensity in a warming world. The consequences are, for instance, higher mortality, species extinction, water scarcity, as well as extreme rates of glacier loss, sea ice loss and sea level rise (IPCC 2023). Another prominent impact example at mid-latitudes is the indisputable suffering of forests under heat stress and droughts (Spiecker and Kahle 2023). This is evident in Central Europe, where increasingly hot and dry summer conditions in the recent decades have caused substantial damage to forests (Buermann et al. 2014; Kolář et al. 2017; Colangelo et al. 2018; Schuldt et al. 2020; Bastos et al. 2021).

In Central Europe, a multitude of dendroecological studies have demonstrated the connection between tree growth and climatic variables, where tree ring widths have served as a crucial metric (Friedrichs et al. 2009; Kraus et al. 2016; Dulamsuren et al. 2017; Debel et al. 2021). Given the evidence of increasing drought stress on forests in recent decades, which also affects the region of Bavaria in southern Germany (Debel et al. 2021), there is a strong need to assess how forest health will continue to change with regard to future climates. The transdisciplinary project BayTreeNet (https://baytreenet.de/) addresses the topic of climate-driven forest stress in Bavaria from the perspectives of three disciplines: climatology, dendroecology, and educational research. Collaborating under a common objective, the wider aim is to enhance public awareness and educate high school students more deeply about the challenges faced by forest ecosystems. A recent publication provides a comprehensive description of the project's conceptual viewpoint (Bräuning et al. 2022), while a first case study for one week in summer 2021 synthesized the three different perspectives (Mölg et al. 2024). The key concept of BayTreeNet is to examine forest stress at the synoptic scale of weather regimes (WRs), adding value to the traditional approach where more aggregated climate variables (e.g., monthly means) have been related to tree dynamics (Deslauriers et al. 2003; Güney et al. 2020). The major goal in the climatology part is to identify WRs in the past and future through an objective approach, which requires a number of steps: choosing an automated classification method, evaluating its performance for the observed climate of the recent past, incorporating global general circulation models (GCMs) that provide data of simulated future climates, and analyze in detail what changes in WRs are likely to occur. In the present paper we target the first three points, which means we test a classification scheme on observation-based climate data and ask the question whether the derived WRs are represented reasonably in specific GCMs. Based on our results, subsequent research can examine how Bavarian forests at different locations have and will respond to the different WRs in possible future climates and scenarios.

In the mid-latitudes (e.g., Central Europe), high- and low-pressure systems are the predominant features defining weather and climate. This has led to the concept of WRs, which are a process-based attempt to categorize the different states of atmospheric large-scale circulation. As many near surface climate variables, including temperature, precipitation, and wind, are highly impacted by these circulation patterns (Riediger and Gratzki 2014), they serve a broad range of purposes within the field of synoptic climatology (Huth et al. 2008). Studies have shown the impact of WRs on health, extreme events, hydrological and ecological processes as well as on many other phenomena (Bissolli and Dittmann 2001; Post et al. 2002; Clark and Brown 2013; Horton et al. 2015; Psistaki et al. 2020; Rousi et al. 2022; Zong et al. 2022; Thomas et al. 2023), also with a specific focus on Central Europe (Plaut and Simonnet 2001; Hertig and Jacobeit 2014; Riediger and Gratzki 2014; Herrera-Lormendez et al. 2022).

WRs have been of interest since 1881 and became well-known when Hess and Brezowsky (1952, 1969, 1977) defined 29 so-called "Großwetterlagen" for the European and North Atlantic regions via a manual classification method (James 2007). These records were kept up to date (Werner and Gerstengarbe 2010) and are continued nowadays by the German Meteorological Service (DWD). However, this is time-consuming and partly subjective due to manual classification.

In order to ensure an objective and automated WR classification that can deal with large multidimensional data, Self-Organizing Maps (SOMs) are employed in this study. SOMs have been demonstrated to be a valuable and powerful tool in climatology and synoptic meteorology for detecting and clustering large-scale meteorological patterns (Hewitson and Crane 2002; Skific and Francis 2012). The SOM is fed meteorological input data that describe the predominant weather dynamics (in our study, sea level pressure (SLP) and wind in the upper troposphere), which it organizes into classes (nodes) that represent different WRs or variants of them. However, not every node corresponds to a WR in the classical "Großwetterlagen" sense of weather patterns, meaning that a single classical weather pattern may encompass multiple nodes. An overview of studies, which demonstrate that SOMs are a skillful and effective method for research questions on WRs, is given in Loikith et al. (2017).

Future projections of climate rely considerably on the selection of realizations from the available GCMs. Recent studies demonstrated that analyses with a small preselected ensemble can lead to more accurate and more robust results for regional case studies (Cassano et al. 2007; Hall et al. 2019; Mölg and Pickler 2022). For this purpose, a new GCM selection technique was recently developed (Pickler and Mölg 2021), which results in a ranking of the most suitable and representable GCMs for a customized region of interest. We apply this technique to Western/Central Europe. Thus, compared to conventional approaches using the full or a large GCM ensemble, in our study the WRs obtained from GCM data are based on carefully preselected models.

The established SOM method is applied to these preselected GCMs in order to evaluate how well GCMs can reproduce observationally-based WRs. Hence, the methodological refinement in this study is the novel approach of combining SOMs with an advanced GCM selection technique. Our main goal is to analyze the characteristics of European summer WRs in the recent past and, thereby, to identify GCMs that can be considered in ensuing research on WRs in future climates. The paper is structured as follows. After an overview of the employed data and the applied methods, we present WR results for the summertime period of 1990-2019 based on ERA5 reanalysis data (see Section 3a). Then, we analyze the simulated WRs in the preselected GCMs for the same period to assess the realism of state-of-the art climate models in reproducing European WRs (see Section 3b). The final section presents the most important conclusions.

## 2. Data and Methods

Due to the strong seasonality in mid-latitude climate, our investigation was carried out on a seasonal basis, consistent with the prevailing approach in automated classification methods like SOMs (Horton et al. 2015; Beck et al. 2016; Gibson et al. 2017; Loikith et al. 2017). The focus is on summer (JJA; 92 days per year), since, in the mid-latitudes, this season is most likely to have the greatest impact on humanity and forest ecosystems regarding the challenges posed by climate change (Hansen et al. 2012).

As input variables for the SOM we used SLP and wind from the upper troposphere (U- and V-wind at 200 hPa for reanalysis and 250 hPa for GCMs due to data availability of daily resolution), which are key variables for the classification of weather patterns (e.g., James 2007; Loikith et al. 2017; Werner and Gerstengarbe 2010). The study region used to define the WRs covers Europe and the northern Atlantic region (30°N-70°N, 50°W-40°E) (Figure 1), which is a common domain for European weather pattern analysis (James 2007). For the climatological analysis of surface air temperature and precipitation, we focus on the larger region of central Europe (45°N-57°N and 0°E-20°E) (Figure 1b). This areal extent is also used for grouping the WRs into main flow types (see Section 2e). The analysis of the recent past covers the period from 1990 to 2019. This period was chosen to fully cover the most recent decade and to have a length of 30 years.

## 2.1 Reanalysis

For the WR analysis, SLP, U-wind at 200 hPa (U200) and V-wind at 200 hPa (V200) data are retrieved from the ERA5 reanalysis dataset produced by the European Centre for Medium-Range Weather Forecasts (ECMWF) (Hersbach et al. 2020). It has a global coverage of land and ocean with a high spatial resolution of 0.25° × 0.25°. The dataset is commonly used in climate science and is also established for studies of atmospheric dynamics (e.g., van der Wiel et al. 2019; Montoya Duque et al. 2021; Lhotka and Kyselý 2022). Moreover, it has already been used for the classification of circulation regimes through SOMs (Rousi et al. 2022; Thomas et al. 2023). Studies have also shown that WR classifications resulting from a SOM are hardly sensitive to the choice of reanalysis (Gibson et al. 2017; Mattingly et al. 2018).

## 2.2 CMIP6

GCM data come from the Coupled Model Intercomparison Project, phase 6 (CMIP6) (Eyring et al. 2016; O'Neill et al. 2016). There are historical and future simulations with different scenarios describing a variety of possible socioeconomic pathways (SSPs) that are detailed in Gidden et al. (2019). Since the GCM historical simulations end in 2014, they were extended with the first five years from the respective SSP1-2.6 projection (Ciavarella et al. 2021) to cover the study period of 1990-2019.

It is in the nature of SOMs that the nodes may be placed in different positions in the SOM array, even when a classification calculation is repeated using the same settings (Reusch et al. 2005). Thus, averaging single nodes across all GCMs like in a traditional ensemble method is misleading and not practical in this study. Also, if the input variables were averaged before the SOM, the data would be smoothed and the pronounced high- and low-pressure systems would not appear. GCM data are therefore mostly presented model by model in the results sections.

## 2.3 GCM selection

The GCM selection technique of Pickler and Mölg (2021) evaluates the ability of GCMs to reproduce the mean state of the climate and the space-time climate variability for the atmospheric key variables in a specific region. Therefore, all available GCMs were culled based on the availability of monthly temperature, specific humidity, U- and V-wind for the historical simulation and SSP1-2.6. To ensure the possibility of using the models in studies on future WRs derived from GCMs, the SSP5-8.5 scenario must also be available. In addition, we only considered models that provide a daily resolution for the historical simulation, SSP1-2.6 and SSP5-8.5, to enable a wide range of possible GCM applications in subsequent research that follow the WR approach of this paper. This left us with 107 realizations from a total of 24 models.

Following the Pickler and Mölg (2021) method, we computed the non-degenerate empirical orthogonal functions (EOFs) of the four atmospheric variables (specified above) at 200 hPa, 500 hPa, and 850 hPa seasonally over the European region (30°N-70°N, 50°W-40°E) from 1990 to 2014 (the end year of the historical simulations in CMIP6) using ERA5 data. Subsequently, we assessed the culled CMIP6 realizations' fidelity in simulating these EOFs, examining variance fraction and shape, totaling 24 assessments. Additionally, we evaluated CMIP6 realizations in simulating the mean annual cycle and the statistical distributions with respect to Bavarian weather station data (e.g., Collier and Mölg 2020) for 2 m air temperature (t2) and precipitation (pr). The differences in the absolute mean, Pearson correlation coefficient for annual cycles, root mean square error (RMSE) for cycle magnitude, and the two-sample Kolmogorov-Smirnov statistic for distribution similarity were considered. For details, refer to Pickler and Mölg (2021).

The resulting "test heatmap" of the Pickler and Mölg (2021) method is presented in the Supporting Information Figure S1. In the ranking, some models (e.g. MPI) are represented several times with a different realization. This means that the same model was run using different initial/starting conditions. Since these realizations only differ in internal variability, we consider one realization (the best-ranked) from a certain GCM if there are multiple realizations available (Jain et al. 2023). Furthermore, we only consider the ranks in the top quarter (rank 1 to rank 26) of the ranking list (Figure S1). From the pool of the remaining models, we finally consider only those models that provide all the required data for the SOM (Table 1). Three models fulfill all criteria and are therefore used in this study: rank 1 (MPI-ESM1-2-LR r29i1p1f1), rank 4 (CanESM5 r1i1p1f1), and rank 12 (MRI-ESM2-0 r5i1p1f1), hereafter referred to as r1, r4, and r12. By using only a few, carefully preselected GCMs we aim to enhance the accuracy of projections in comparison to a common large GCM ensemble, following studies that highlight the value of a small, preselected ensemble for regional case studies (Cassano et al. 2007; Hall et al. 2019; Mölg and Pickler 2022).

## 2.4 Self-Organizing Maps

There are different classification methods, subjective and objective ones, for circulation types and synoptic dynamics (Huth et al. 2008; Beck and Philipp 2010; Philipp et al. 2010; Lewis and Keim 2015; Alvarez-Castro et al. 2018; Hidalgo and Jougla 2018; Stryhal and Huth 2019). Naturally, there is no single best method. However, objective and automated classification systems have proven more suitable for most research questions like ours (Philipp et al. 2010).

In the present study, the classification of WRs was calculated using SOMs, also known as Kohonen Maps (Kohonen et al. 2001). In the field of machine-learning, SOMs are classified as an unsupervised learning technique. They organize the input data (here, SLP and winds in the upper troposphere) into a user-defined number of classes (nodes), which are created by iteratively adapting the nodes to the input data (Loikith et al. 2017). SOMs are known to be a useful, robust, objective and time efficient tool for studying synoptic-scale meteorology in observations and climate models (Hewitson and Crane 2002; Reusch et al. 2005; Skific et al. 2009; Sheridan and Lee 2011). Another benefit is the arrangement of the individual classes (nodes) on the map, which ensures that similar patterns are positioned nearby, while strongly different patterns are separated by greater distances (Hewitson and Crane 2002). The cited studies offer comprehensive descriptions of the SOM methodology, analyses and performance assessments, hence this paper only provides a concise overview that is relevant to the current research.

The input variable SLP has often been applied to SOMs in the context of WR classification (e.g., Reusch et al. 2005; Cassano et al. 2007; Skific and Francis 2012; Loikith et al. 2017; Thomas et al. 2023). Classical WRs are often divided into zonal or meridional types (e.g., Bissolli and Dittmann 2001), therefore the U- and V-wind components in the upper troposphere are also included as input for the SOM. Further variables were tested individually (wind speed at 200 hPa (Loikith et al. 2017), only U-wind and only V-wind) as possible input variables, but they did not enhance the results. The geopotential height at 500 hPa is also a frequently used variable in studies to identify and classify WRs (e.g., James 2007). However, we have deliberately decided against this variable because it is not suitable for detecting and investigating future WRs (which follow-up research will do) due to the potential effects of thermal expansion of the troposphere, which is expected to lead to a generally higher geopotential height (Horton et al. 2015). Also, the variables analyzed had a daily resolution due to the typical daily variability of WRs. Following Loikith et al. (2017) we carried out a pre-processing of input data before calculating the SOM: each grid point's data were first normalized by the temporal standard deviation (for JJA) and then weighted by the square root of the cosine of latitude (Johnson et al. 2008).

The most essential decision for creating a SOM concerns the number of nodes and the array size (Reusch et al. 2005), with the choice of node number being guided by the specific research question. There must be a balance between representing a reasonably complete range of major patterns and the feasibility to interpret the results. Too many nodes could lead to a high dispersion of information resulting in the inability to draw conclusions or recognize correlations. Too few nodes, on the other hand, could cause an overly strong generalization (Loikith et al. 2017). Previous studies have tested and used a range of different node numbers and set ups. The most typical array sizes varied between 2×3 and 6×7 (Cassano et al. 2015; Cassano et al. 2016; Gibson et al. 2017; Thomas et al. 2023), considering that no SOM with a square setup should be chosen because of mathematical issues (Reusch et al. 2005). To evaluate the quality of the classification for different SOM sizes, the distribution of Pearson correlation and Root Mean Square Error between the observed variable at each day and the winning node pattern of that day can be calculated, following Gibson et al. (2017). We quantitively and qualitatively evaluated the results obtained with different SOM arrays (9, 12, 20, 30 nodes). As anticipated, the inclusion of more nodes had a slightly positive impact on the arithmetic mean of the correlation, as this increase in nodes corresponds to a greater level of detail in the representation of circulation patterns (Johnson et al. 2008). In conclusion, we found that a 4×3 SOM is most suitable for our research question, resulting in twelve nodes that represent twelve WRs or variants of them. This array size is capable of capturing the range of synoptic-scale variability with enough detail to distinguish between different variants of the same pattern, while allowing physical interpretations. The size is in line with other studies (Loikith et al. 2017; Thomas et al. 2023).

In addition to tests for the array size of the SOM, we did several sensitivity tests and tried different parameter settings in the SOM calculation: topology (rectangular and hexagonal), number of iterations (100 and 1000), search radius (1 and 4) and learning rate (0.05 and 0.1). The differences in the results using the various settings were generally negligible or non-existent, which is in line with other studies (e.g., Gibson et al. 2017). In the end we conducted the analysis with the "optimal" SOM setting shown in Table 1, based on the maximum coefficient of the Pearson correlation between the observed variable at each day and the winning node pattern of that day (Johnson et al. 2008; Gibson et al. 2017) and the lowest RMSE.

## 2.5 Grouping into main flow types (large weather types)

In order to simplify the interpretation and evaluation of the results and to better compare the different datasets (reanalysis and GCMs), the 12 resulting nodes of the SOM for each dataset were organized into three groups, so-called 'large weather types' (LWTs). These LWTs represent the main flow types which depend on the main flow direction: zonal (ZON), meridional (MER) and mixed (MIX), similar to the classic European "Großwetterlagen" (Werner and Gerstengarbe 2010). Categorizing or grouping weather patterns into a few main groups of LWTs is a common approach when investigating large-scale circulation patterns (e.g., Alvarez-Castro et al. 2018; Messori and Dorrington 2023). We developed an automatic grouping to achieve an objective and reproducible result, based again on our specified region of larger central Europe (45°N-57°N and 0°E-20°E). We differentiated between the three LWTs by calculating the zonal-to-meridional-ratio of wind direction (ZMR) of each node. For this purpose, the wind direction was calculated from U200 and V200 for each grid point and node, and the number of grid points falling within a 90° sector around the four main wind directions was counted, similar to Dittmann (1995). To eventually obtain the ZMR value for each node, the zonal component (East plus West) is divided by the meridional component (North plus South). If a node does not have a meridional component, it is divided by 0 and the result is infinite.

The following threshold values were determined using the ERA5 dataset (Table 2): ZMR = infinite (ZON), ZMR ≤ 3 (MER) and ZMR > 3 (MIX). A dominance of the zonal component is climatologically determined due to the location in the mid-latitudes under the strong influence of the westerlies (see also later in Section 3a). Therefore, the threshold of ZMR > 3 for meridional patterns is above 1 because the mean background flow is westerly.

To demonstrate that ZMR = 3 is a suitable threshold, the histograms for all nodes are provided in the Supporting Information (Figure S2), which show that the next highest ZMR-value, after 3, is nearly twice as high (ZMR = 5.83; N09). This substantial difference indicates a statistical differentiation between the LWTs and the choice of 3 as a threshold to be practical. Also, by using 3 it is obvious that MER types clearly differ from the climatological distribution (Figure S2).

## 3. Results and Discussion

In this section, the detected summer WRs derived from the ERA5-driven SOMs are investigated for their frequency, their main flow type and their relation to near-surface temperature and precipitation in the last three decades (1990-2019) (Section 3a). Afterwards the simulated WRs derived from the three preselected GCMs are evaluated and analyzed for the same period (Section 3b).

### 3.1 Weather regime characteristics (ERA5 1990-2019)

Each node possesses distinguishable characteristics which allow us to recognize certain circulation types. Note that the WRs derived in this study are not comparable to classical weather patterns, such as those employed by the DWD, as the WRs in this study result from an automated classification and capture mean conditions across multiple days, rather than individual days that are considered in manual classifications. Figure 1 shows the composite means for SLP and wind at 200 hPa (W200) of the resultant twelve nodes for summer, referred to as node 1 (N01), node 2 (N02), etc. The individual nodes occur with probabilities ranging from 4.8% to 10.8% shown in the title, a range commonly observed in large-scale weather pattern frequencies obtained through SOMs, signifying a relatively uniform distribution (Loikith et al. 2017). Given that the SOM places similar nodes together, it is evident that patterns with less pronounced pressure centers are located on the lower left side of the SOM. All regimes dominated by an extensive high-pressure system (anticyclones) are in the right region of the map (N03, N04, N08 and N12). The pattern with the most well-defined Icelandic low and Azores high can be observed at N08 (9.6%). The three rarest nodes (N05, N07, N11) do not occur every year. They show a high-pressure system far to the North (N07, N11) or a low-pressure system far to the south (N05), indicating pronounced wave structures, which are not standard conditions, and which also appear in the wind field (W200) in the upper troposphere (Figure 1b). Regarding the latter, the predominance of the westerlies is obvious throughout the nodes.

As described in Section 3e, the nodes were grouped into the three LWTs that can also be localized fairly well in the SOM array: ZON (upper right area of the SOM), MER (lower right area) and MIX (left column and N04) (Table 2). Figure 2 illustrates an example of each LWT with the frequency distribution of the wind direction. It can be clearly seen that zonal nodes (ZON; ZMR = Inf) have a pronounced peak in the West sector (Figure S2) and have no information in the North or South sector. In addition, these nodes are very similar to the mean summer wind conditions in 1990-2019 (grey shading). Meridional nodes (MER; ZMR ≤ 3), on the other hand, show a substantial amount of northerly or southerly flow and mostly have a peak in one of these two sectors. Mixed nodes (MIX; ZMR > 3) show the peak in the West sector that reflects the west wind zone, but also have additional northerly and/or southerly winds. The three LWTs occur with probabilities of 34% (ZON), 31% (MER) and 35% (MIX). This distribution agrees well with the summer average from 1881-2008 of the large weather types as reported by the DWD (Werner and Gerstengarbe 2010), see Figure 3 (it also includes data to be discussed later in Section 3b).

Next, we look at the climatological characteristics of t2 and pr for each node in our defined Central European region by calculating the composite mean of the variables. Spatial anomalies with respect to the reference period 1990-2019 (Figure 4) and the associated timeseries were calculated for each node (Figure S3). No large spatial t2 anomaly variability was detected within a node apart from N01 (N12) showing a light transition from eastern (southern) negative to western (northern) positive anomalies (Figure 4a). N09 can be clearly identified as the warmest node with anomalies of up to +4°C (relative to the mean of 15.4°C). In addition, N09 displays the lowest average daily rainfall of 1.3 mm (Figure 4b and Figure S3b), compared to a mean value of 2.8 mm per day considering the whole period and every node. N07 is the coldest node with a mean temperature of 15.3°C and an anomaly of up to -4°C. Looking at the spatial differences of the pr anomalies within individual nodes (Figure 4b), the following can be observed: N11 (and N12) have wetter conditions South of the Alps probably as a result of a drained low-pressure system over the Mediterranean, which pushes precipitation from the Mediterranean towards the southern Alps. N09 (N10) shows drier conditions in the east (west), as the ridge is located further east (west) (Figure 1b).

To try to summarize all important aspects, the "heatmap" in Figure 5 attempts a synopsis of the frequency, duration, and intensity (t2 and pr) characteristics for each node. Significances of temporal trends are based on a linear regression. N03 stands out with the only significant frequency trend of about +2% per decade and also shows a significant increase for pr, indicating more frequent wetter days over time. In terms of duration, N05, N06 and N07 are noticeable, lasting 3.1 days on average. Regarding t2 and pr anomalies over the whole record (1990-2019), the most important outcome is that the data portray a consistent positive trend of t2 across all nodes except for N05. Five of these warming nodes show a statistically significant warming trend, which is a clear implication of global warming at the European synoptic scale. Not only warm WRs have intensified, cold WRs show a significant increase in t2 as well (N04, N10). No clear patterns are recognizable for pr anomalies. Regarding the development over time, three out of four significant pr trends have a positive sign (N02, N03, N12), but with a relatively small average magnitude of +35 mm per season and decade.

## 3.2 GCM WR performance (1990-2019)

The application to ERA5 reanalysis data demonstrates the capability of SOMs to identify realistic WRs in central Europe. Thus, the approach can now be applied to GCM data. Model realism is validated for the period of 1990-2019 by comparing the top three GCMs (see Section 2c) to the ERA5 data in terms of (i) the results of the SOM (frequency and duration of the nodes, i.e. WRs), (ii) the LWT frequency distribution, (iii) SLP and wind fields of the nodes and (iv) the climatological characteristics of the nodes (heatmap).

The density plots (Figure 6) show that the models successfully capture the statistical characteristics of the nodes, as both the frequency and the duration of the nodes are similar across all datasets and show an overlap with the ERA5 data. Especially for the node frequency, the range is reproduced well, with no shift of the GCM data to the left or the right (Figure 6a). Regarding duration, slightly longer-lasting WRs are observed in the GCMs compared to the reanalysis (Figure 6b). The average duration of a node varies between 2.7 days (ERA5) and 3.2 days (r4) and is nearly 3 days across all datasets.

To assess the realism of the individual nodes and their synoptic features, the grouping into the three LWTs based on wind direction was applied to the GCMs. The frequencies of the individual types show an overall agreement with the reanalysis data (Figure 3). One GCM (r1) shows a dominance of ZON WRs and fewer MIX WRs. This discrepancy can likely be attributed to the positive zonal bias in the mean upper-level winds over the North Atlantic-European region observed in the MPI-ESM-1.2-LR model (r1) (Müller et al. 2018). Overall, it can be concluded that the reanalysis results are consistently within the range of variability of the model ensemble, indicating that the models capture the essential characteristics of the reanalysis data effectively.

This is further illustrated below using the best-ranked GCM (r1) as an example. One node from each LWT group was selected to demonstrate that the large-scale meteorological fields in the models are realistic (Figure 7). Note that the grouping into the LWTs is based on wind direction, so the agreement in SLP patterns shown here underlines the quality of the results even further. The zonal pattern shows a pronounced high-pressure system over the Azores, the meridional pattern shows a high-pressure bridge between the Azores and Great Britain, and the mixed pattern is characterized by a diminished Azores High and a North Atlantic low-pressure zone that pushes into Northwest and Northern Europe. It should also be noted that the mean frequency of the respective 'pairs' is very similar with a difference of only 0.9% - 1.3% per summer season. In summary, the SOM nodes generated from GCM data display the major synoptic circulation patterns obtained from the ERA5 data in European summer, and the selected GCMs can represent the WRs realistically.

To examine the statistical and climatological characteristics of the nodes in a compact way, a heatmap was also created for each GCM dataset (analogue to Figure 5 for the ERA5 data). To simplify the message and reduce the dimensions, the results of the four heatmaps were summarized in a bar chart (Figure 8) to enable a better comparison of the fundamental properties. This figure focuses on the number and direction (positive or negative) of significant temporal trends. Also, it illustrates the sign of the anomaly in near-surface atmospheric conditions (t2 and pr) as + and – symbols. Frequency and duration (left side of Figure 8) together show very few significant trends in both directions, ranging between 0 to 3 per dataset. Slightly more significant trends can be observed for pr, most of which have a positive seasonal anomaly, also in the range of 0 to 3 per dataset. However, t2 shows a relatively high number of significant changes over the last three decades, all with a positive trend and affecting up to 10 out of 12 nodes per dataset. Furthermore, the anomaly signs (plus and minus shown on the bars) of the nodes with significant trends show agreement between the reanalysis and the GCMs, as the temperature rise trends contain nodes with both negative and positive anomalies. In summary, the most important finding from Figure 8 is that the ERA5 results of all four categories lie within the range of the models, which suggests that the small GCM ensemble is able to capture the observed conditions.

All results shown in this section indicate GCM skill in representing the circulation variability realistically, implying that the selected GCMs are a useful data source to study future summer WRs in Central Europe. We therefore recommend the usage of these three GCMs for climate projections in Central Europe in the frame of WR-related research. Similarly, previous studies (e.g., Cassano 2007) have conducted their regional future projections using only a small, selected ensemble of GCMs, highlighting the approach of relying on a limited but representative set of models.

## 4. Conclusion

The utility of SOMs for analyzing large-scale synoptic WRs based on daily SLP and winds in the upper troposphere from reanalysis and GCM data has been demonstrated in this study for the Central European summers. We analyzed the frequency and duration of the nodes (i.e., WRs), the resultant major LWTs (ZON, MER and MIX), SLP and wind fields of the nodes, as well as the associated temperature and precipitation characteristics for the recent past (1990-2019). The analysis was conducted using a preselected subset of GCM realizations from an independent method (Pickler and Mölg, 2021) to ensure that the climatological conditions of Central European summers and their variability are sufficiently well represented in those models.

Based on the resulting WRs from the reanalysis, the three major LWTs occur with probabilities of 34% (ZON), 31% (MER), and 35% (MIX), closely matching summer averages of the large weather types reported by the German Weather Service, DWD. The distinguishable features of each node enable the identification of different WRs or variants of them, indicating that the SOM captures predominant European summer synoptic patterns. Regarding the associated temperature and precipitation conditions, the salient result is a positive trend in t2 across nearly all nodes. Roughly half of them show significant warming trends, underscoring the impact of global warming on a European synoptic scale. There is much less consistency between nodes for precipitation trends in 1990-2019.

The evaluation of GCMs against ERA5 data for the period 1990-2019 confirms the models' realism through four key comparisons. (i) Node statistics: Both ERA5 and GCM nodes show very similar frequency and duration characteristics of WRs. (ii) Synoptic circulation patterns: The SLP and upper-tropospheric wind fields of GCM nodes have close analogs in the ERA5 nodes, indicating that the selected GCMs can portray major European summer circulation patterns. (iii) LWT distribution: A more generic classification of WRs into major flow types (LWTs) shows a similar frequency distribution between ERA5 and GCM data. (iv) Climatological trends: As in ERA5, the dominant signal in GCMs is a high number of nodes that show a warming over time, and no strong signal in precipitation trends.

Based on these aspects, the selected GCMs clearly show skill in representing large-scale WRs accurately and, thus, serve as valuable tools for studying synoptic weather patterns during summer in Central Europe. Therefore, the three GCMs – MPI-ESM1-2-LR r29i1p1f1, CanESM5 r1i1p1f1 and MRI-ESM2-0 r5i1p1f1 – seem worthwhile for consideration in studies of WR-related climate projections in this region. Despite some (and expectable) differences in the GCM data, the key is that the ERA5-based results are always within the range of our small GCM ensemble, and that no systematic biases of the three GCMs are evident.

In the context of environmental decision-making, robust future climate projections are essential. The presented combination of SOMs with a recent, novel GCM selection technique allows us to process and visualize large amounts of data more effectively and improve the confidence in certain aspects of future climate estimations, such as future WRs. The findings of this study, therefore, have practical implications since regional weather anomalies strongly depend on large-scale atmospheric circulation as captured in WRs. GCMs also offer a tool and boundary condition to estimate both future regional and local climates through downscaling, by establishing a statistical relationship between the large-scale and regional/local climate conditions (e.g., Tang et al. 2016) or by driving a regional atmospheric model with GCM data (e.g., Teichmann et al. 2013). Furthermore, the analysis of future WRs can reveal links between WRs and regional climate impacts (e.g., heat waves, droughts, wildfires). A next step in this regard will be to analyze the characteristics of WRs in the three selected GCMs up to 2100, which should serve the assessment of the climatological characteristics of future European summers further.

## Acknowledgments

The authors gratefully acknowledge the scientific support and HPC resources provided by the Erlangen National High Performance Computing Center (NHR@FAU) of the Friedrich-Alexander-Universität Erlangen-Nürnberg (FAU). NHR funding is provided by federal and Bavarian state authorities. NHR@FAU hardware is partially funded by the German ResearchFoundation (DFG).

## Conflict of Interest statement

The authors declare no conflicts of interest.

## Funding statement:

This research was funded by the Bavarian State Ministry of Science and Arts, as part of the Bavarian Climate Research Network (bayklif), project "Talking Trees: Schnittstelle von Klimadynamik, Dendroökologie und Bildung für nachhaltige Entwicklung".

## Authors' contributions

**Sibille Wehrmann:** Conceptualization; investigation; writing – original draft; visualization; data curation; methodology; formal analysis; writing – review and editing.

**Carolyne Pickler**: Investigation; writing – review and editing; visualization; data curation; methodology.

**Marlene Schramm**: Investigation; data curation; methodology.

**Thomas Mölg**: Conceptualization; funding acquisition; project administration; writing – review and editing; supervision.

## Data Availability Statement

Publicly available datasets were used in this study. The data can be found at: ERA5 (https://cds.climate.copernicus.eu), CMIP6 (https://esgf-node.llnl.gov/search/cmip6/), DWD (https://opendata.dwd.de/climate_environment/CDC/).

# Tables

*Table 1. Self-Organizing Map set-up used in this study.*

| Parameter | Setting |
|---|---|
| Input variables | daily SLP; daily U200/V200 (ERA5) or U250/V250 (GCMs) |
| Time period | 1990-2019 |
| Area | 30°N-70°N and 50°W-40°E |
| Nodes | 4×3 = 12 |
| Topology | rectangular |
| Number of iterations | 100 |
| Search radius | 1 |
| Learning rate | 0.05 |

*Table 2. Overview of the three large weather types (LWTs) and assigned nodes based on the zonal-to-meridional-ratio (ZMR). Data basis: ERA5 1990-2019.*

| LWT | ZMR | Assigned nodes |
|---|---|---|
| ZON | ZMR = infinite | N02, N03, N07, N08 |
| MER | ZMR ≤ 3 | N06, N10, N11, N12 |
| MIX | ZMR > 3 | N01, N04, N05, N09 |

# Figures

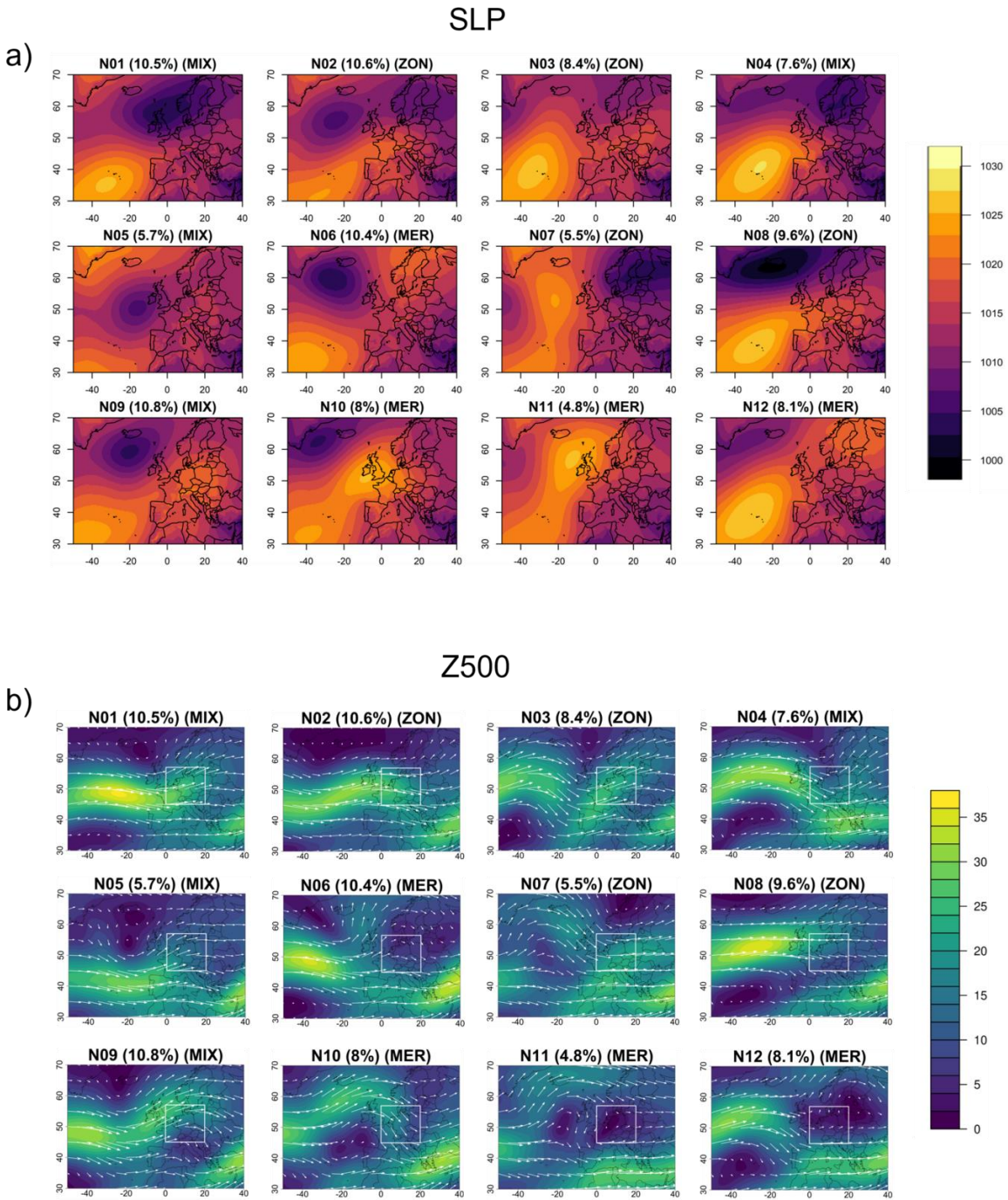

*Fig. 1: JJA Self-Organizing Map (SOM) patterns of a) sea level pressure (SLP in hPa) and b) wind at 200 hPa (W200 in m/s) for each of the twelve nodes over the period 1990-2019. The percentage of days (out of 2760) assigned to each node is shown in the corresponding title. The white rectangle marks the larger central Europe region used for the large weather types (LWTs) grouping and the climatological analysis.*

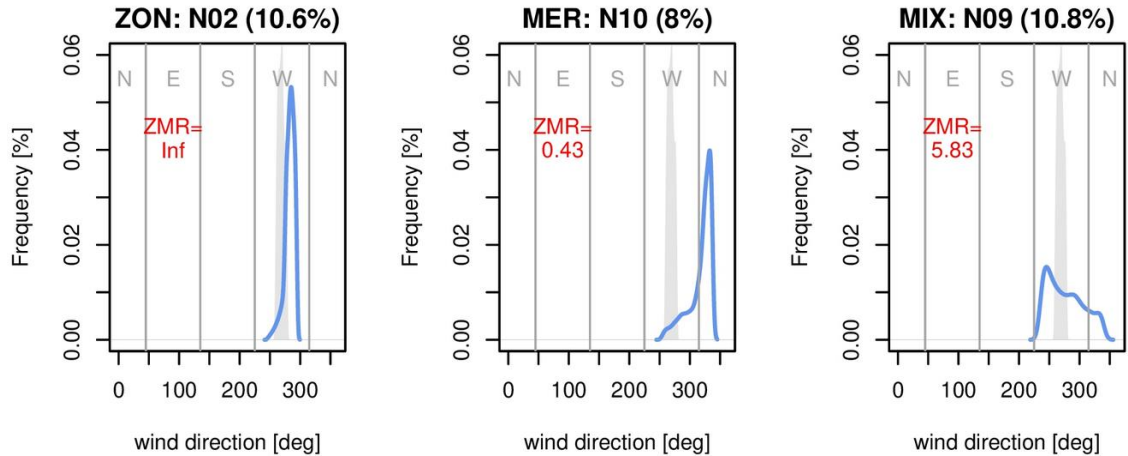

*Fig. 2: JJA histograms of wind directions for three example nodes of the three large weather types (LWTs): zonal (ZON) (N02), meridional (MER) (N10) and mixed (MIX) (N09) for the period 1990-2019. Red number displays the zonal-to-meridional-ratio (ZMR) value of each node. The grey shadow shows the histogram of the mean wind direction distribution in summer for all nodes. Inf= Infinite.*

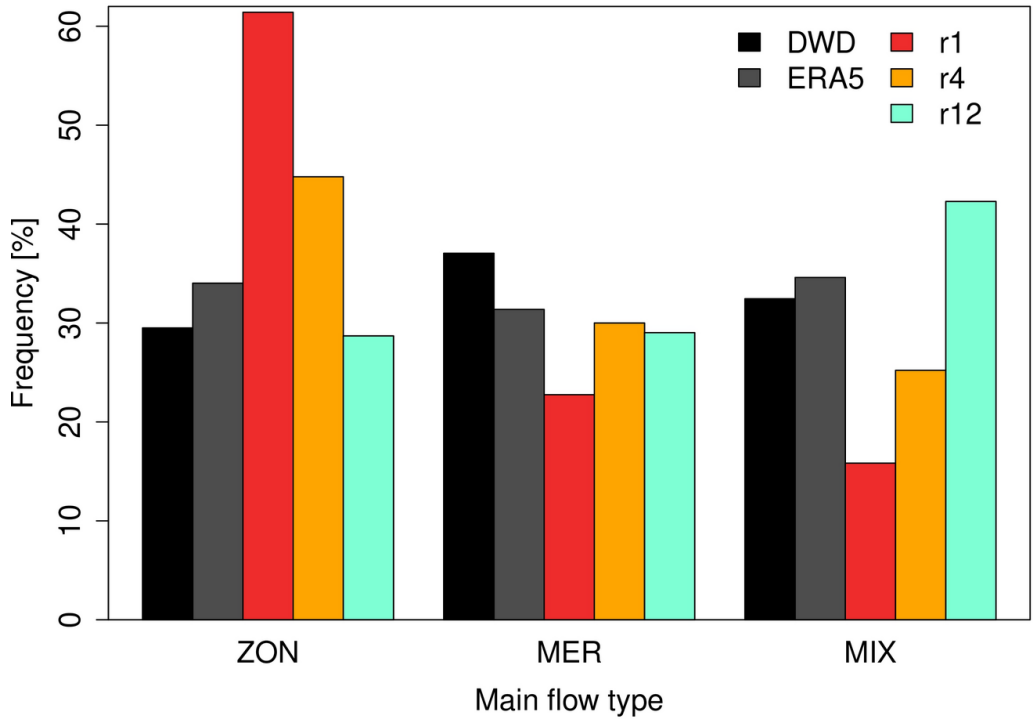

*Fig. 3: Mean JJA frequency of the three large weather types (LWTs) zonal (ZON), meridional (MER) and mixed (MIX) in each dataset for the period 1990-2019. DWD data covers the period 1881-2008 (Werner and Gerstengarbe 2010). r1 = MPI-ESM1-2-LR r29i1p1f1, r4 = CanESM5 r1i1p1f1 and r12 = MRI-ESM2-0 r5i1p1f1.*

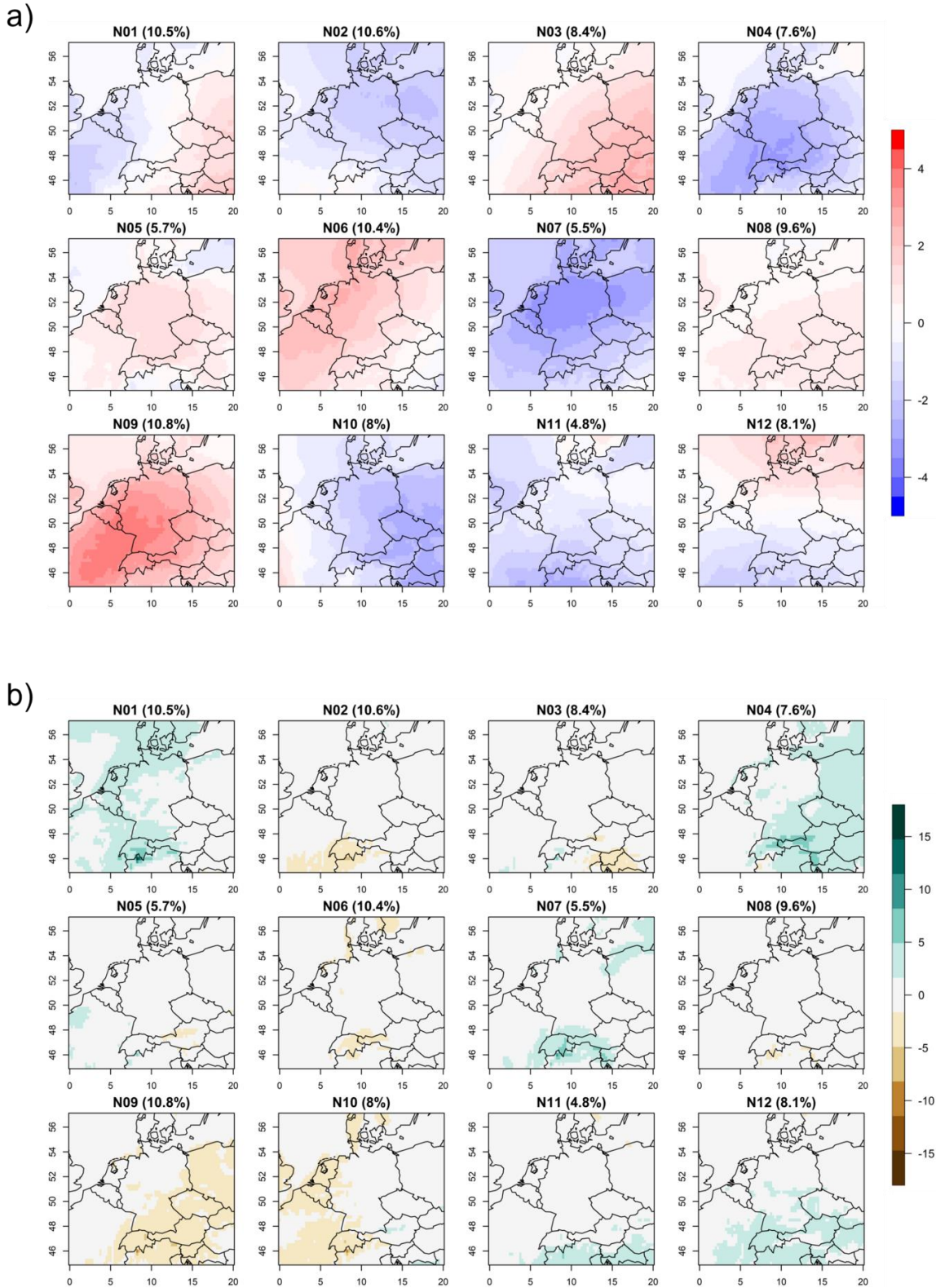

*Fig. 4: a) Composites of daily JJA 2 m air temperature anomalies in °C (reference period 1990-2019) and b) precipitation anomalies in mm for each node for the period 1990-2019. The percentage of each node is shown in the corresponding title. Data basis: ERA5.*

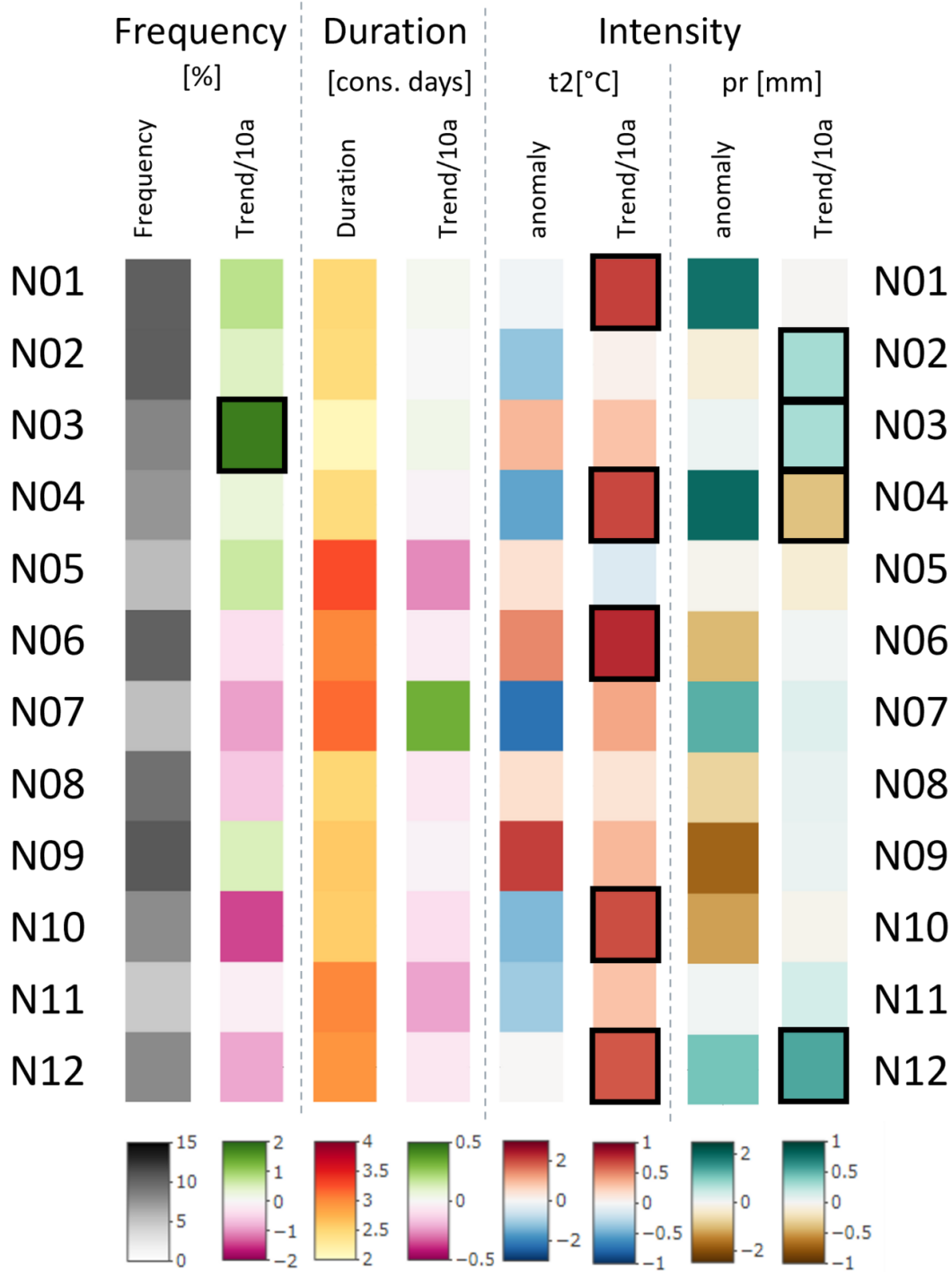

Fig. 5: Heatmap for JJA nodes for the period of 1990-2019 including the summary of the frequency, duration, and intensity (t2 and pr) characteristics for each node. Due to a small sample size, we set the significance level for trends at p<0.1. Significant trends are highlighted with black squares. Anomalies were calculated using the reference period 1990-2019.

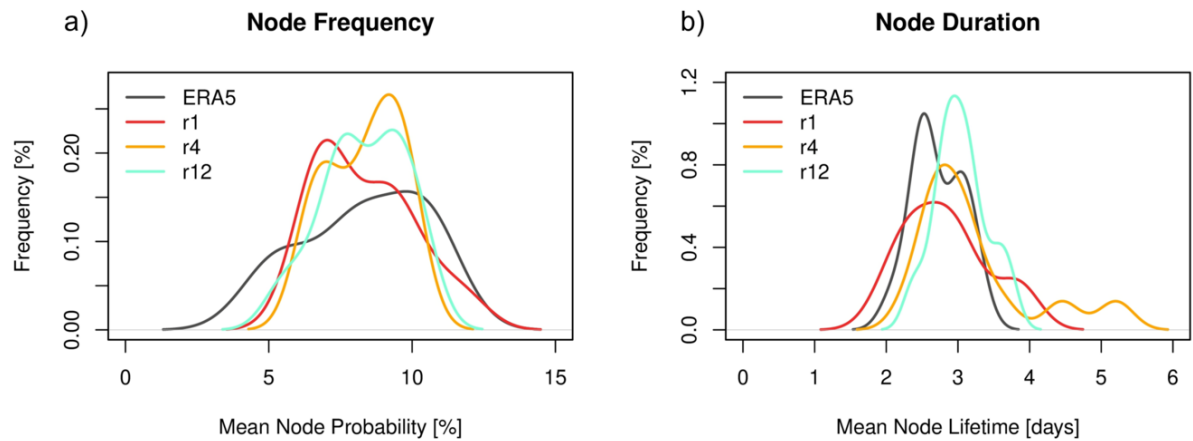

*Fig. 6: JJA density plots of Self-Organizing Map (SOM) results (frequency and duration) for the period 1990-2019 in ERA5 and GCMs. r1 = MPI-ESM1-2-LR r29i1p1f1, r4 = CanESM5 r1i1p1f1 and r12 = MRI-ESM2-0 r5i1p1f1.*

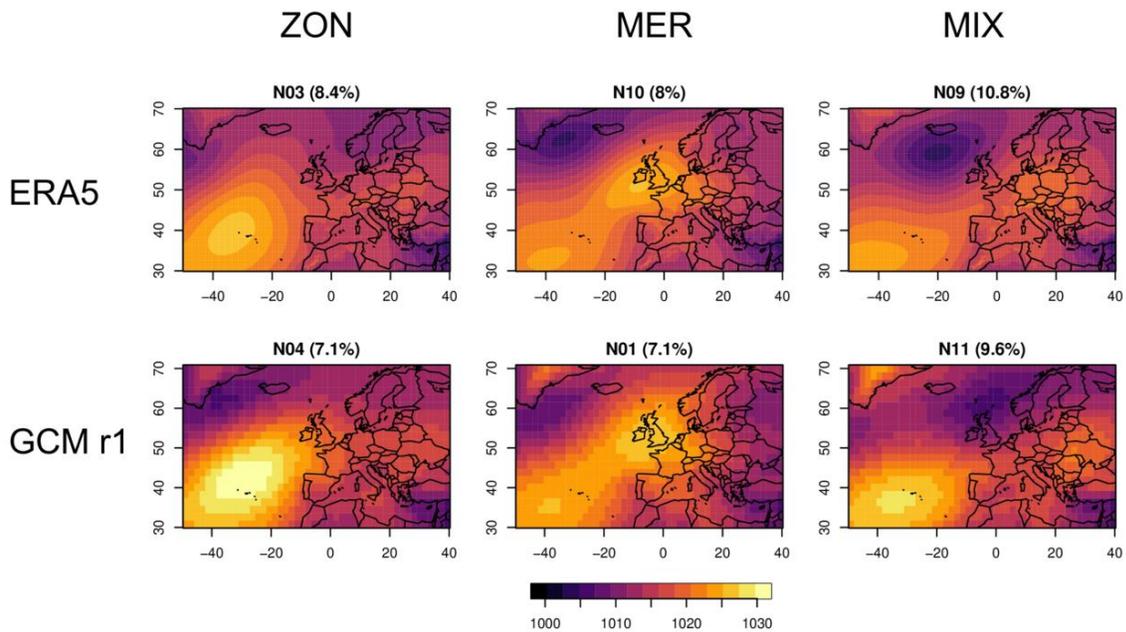

*Fig. 7: JJA Self-Organizing Map (SOM) patterns of sea level pressure (hPa) for each of the large weather types (zonal (ZON), meridional (MER), mixed (MIX)) over the period 1990-2019. Data basis: ERA5 and GCM r1= MPI-ESM1-2-LR r29i1p1f1.*

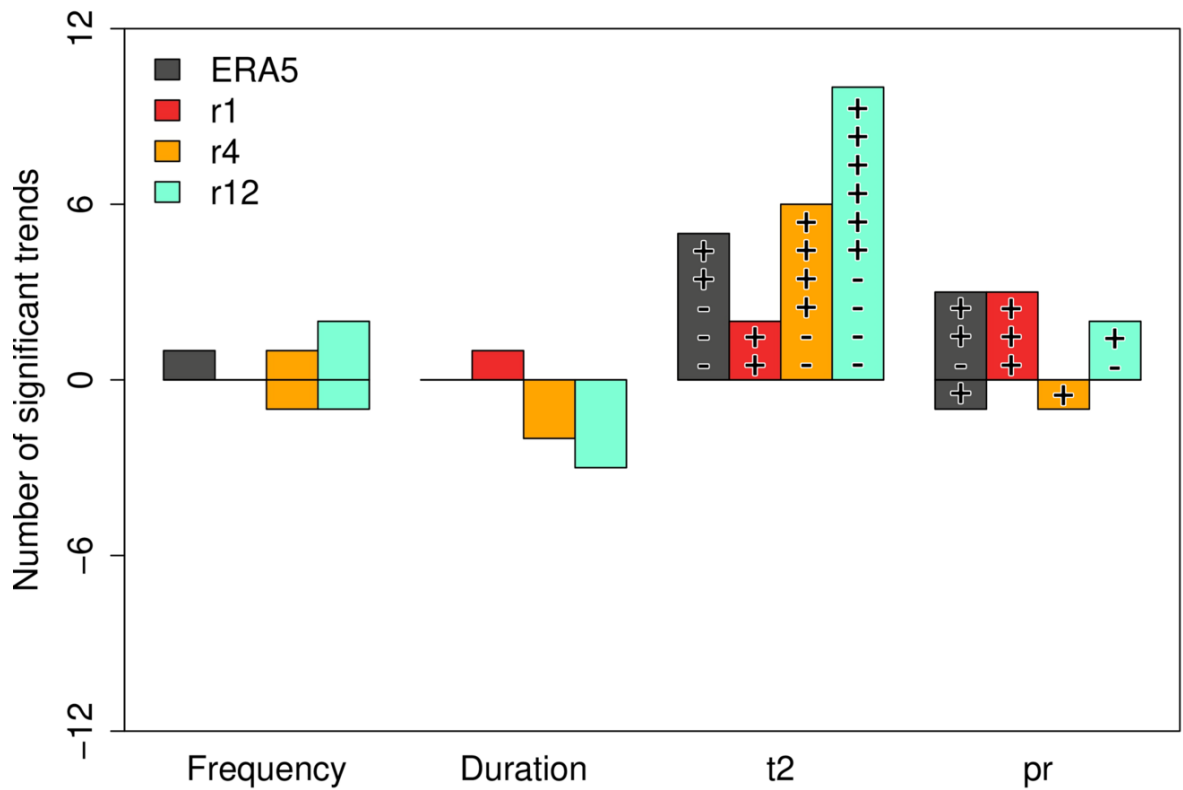

*Fig. 8: Number and direction of significant trends, resulting from the 1990-2019 JJA heatmap, for all four datasets (ERA5 and three GCMs) for the categories of node frequency and duration, and t2 and pr. Negative trends are displayed on the negative y-axis. Y-axis ranges from -12 to 12 to illustrate maximum possible number of significant positive and negative trends. Node anomaly signs are indicated by plus and minus respectively shown on the bars.*
*r1 = MPI-ESM1-2-LR r29i1p1f1, r4 = CanESM5 r1i1p1f1 and r12 = MRI-ESM2-0 r5i1p1f1.*

# Supporting Information

## GCM Ranking List

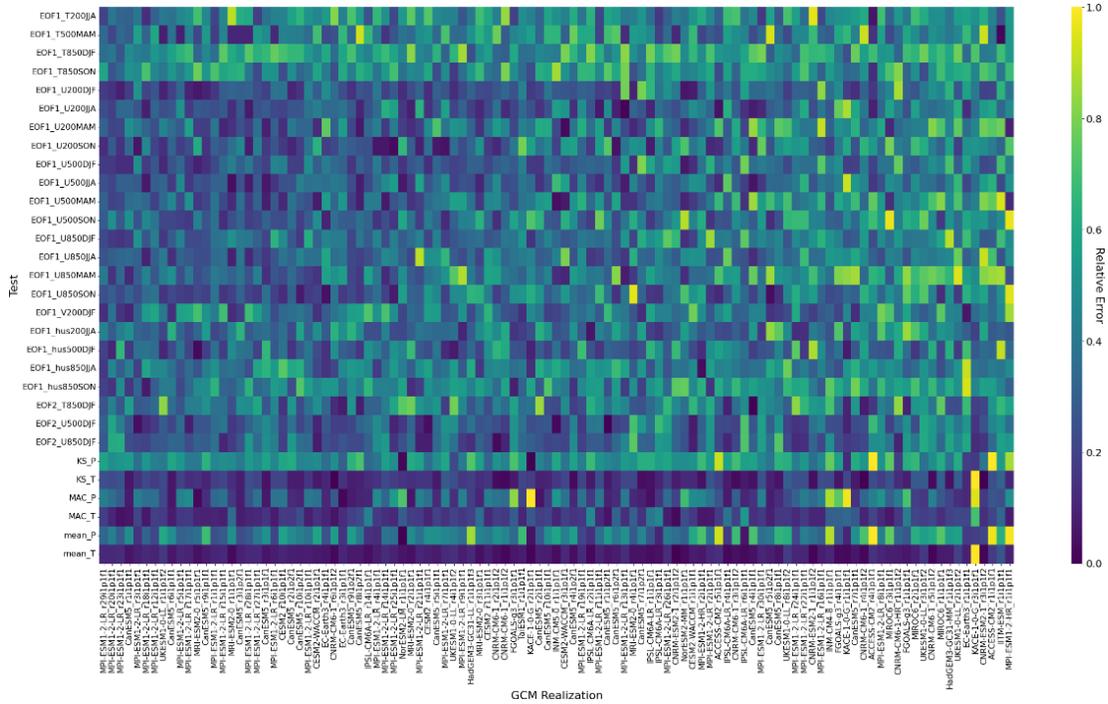

*Fig. S1: Result of the GCM selection technique organized by ranking from left to right. For details of the test procedure, refer to Pickler and Mölg (2021).*

# Histograms of wind direction (ERA5 1990-2019)

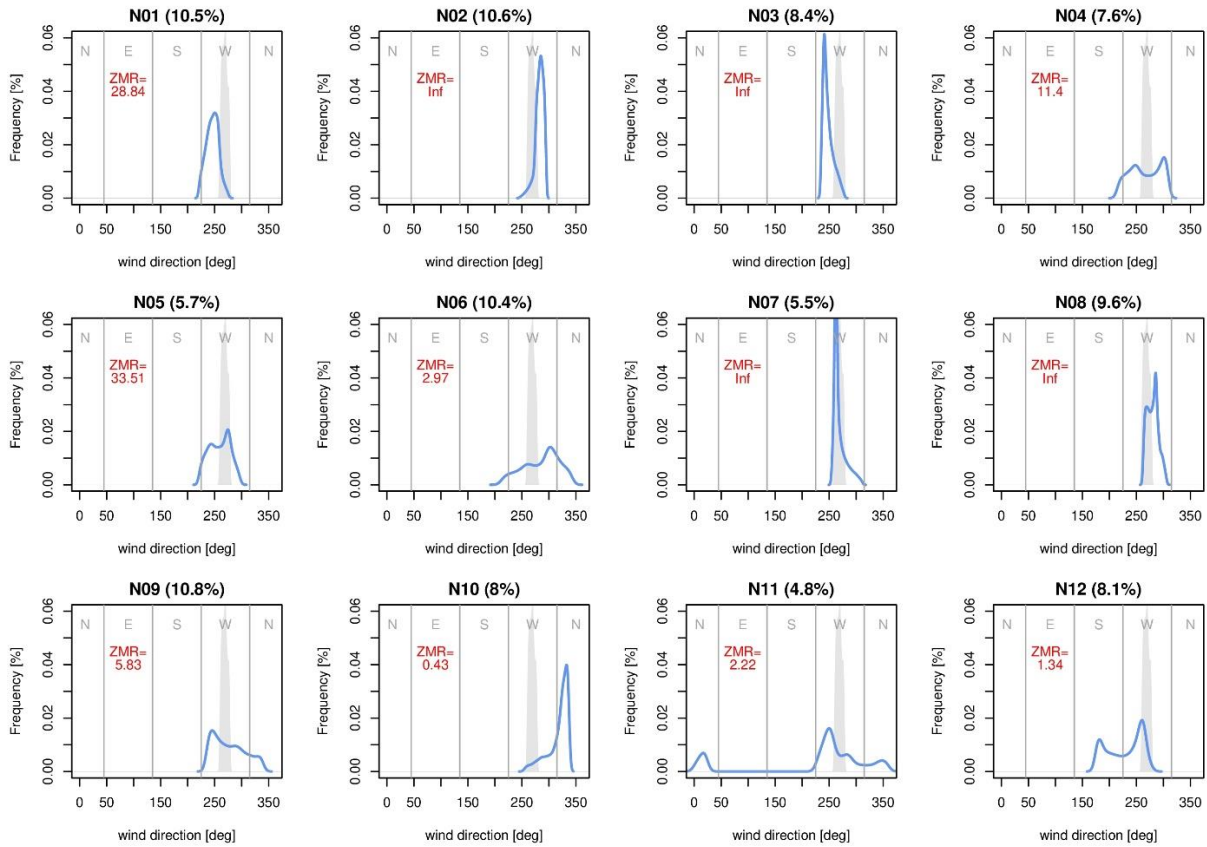

*Fig. S2: JJA histograms of wind directions for all nodes for the period 1990-2019. Red number displays the zonal-to-meridional-ratio (ZMR) value of each node. The grey shadow shows the histogram of the mean wind direction distribution in summer for all nodes. Inf= Infinite.*

**2 m air temperature and precipitation (ERA5 1990-2019)**

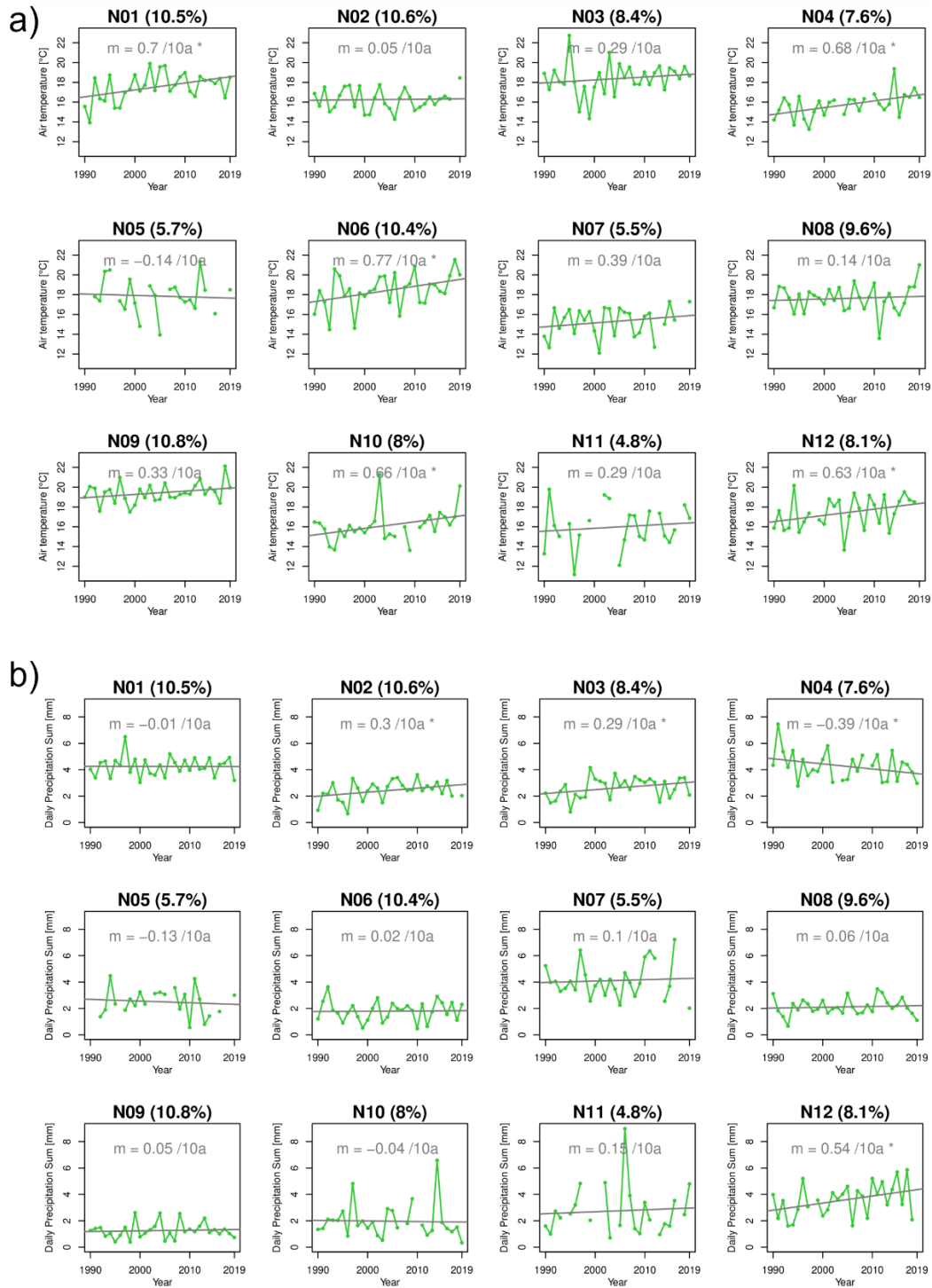

*Fig. S3: Intensities with trends for the period 1990-2019 for a) 2 m air temperature and b) precipitation. Node probability in the title. Significant trends (p<0.1) are marked with an asterisk. The percentage of each node is shown in the corresponding title. The timeseries show missing values when a node did not occur in the respective year. Data basis: ERA5.*